\documentclass[aps,prb,reprint]{revtex4-1}
\usepackage{amsmath}
\usepackage{amssymb}
\usepackage[utf8]{inputenc}
\usepackage{lmodern}

\usepackage[final]{graphicx} 

\usepackage{adjustbox}

\usepackage{braket}
\usepackage{xcolor}

\makeatletter
\DeclareRobustCommand*\textsubscript[1]{%
  \@textsubscript{\selectfont#1}}
\def\@textsubscript#1{%
  {\m@th\ensuremath{_{\mbox{\fontsize\sf@size\z@#1}}}}}
\makeatother

\newcommand{\sub}[1]{\textsubscript{#1}}
\newcommand{\bvec}[1]{\boldsymbol{\mathrm{#1}}}
\newcommand*{\markerfour}{\ensuremath{\vcenter{\hbox{\includegraphics{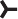}}}}}
\newcommand*{\markerasterisk}{\ensuremath{\vcenter{\hbox{\includegraphics{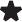}}}}}
\begin{document}


\title{Coupling of shells in a carbon nanotube quantum dot}

\author{M. C. Hels}
\author{T. S. Jespersen}
\author{J. Nyg\aa rd}
\author{K. Grove-Rasmussen}
\affiliation{Center for Quantum Devices and Nano-Science Center, Niels Bohr Institute, University of Copenhagen, Universitetsparken 5, 2100~Copenhagen \O, Denmark}
\date{\today}

\begin{abstract}
    We systematically study the coupling of longitudinal modes (shells) in a carbon nanotube quantum dot.
    Inelastic cotunneling spectroscopy is used to probe the excitation spectrum in parallel, perpendicular and rotating magnetic fields.
    The data is compared to a theoretical model including coupling between shells, induced by atomically sharp disorder in the nanotube.
    The calculated excitation spectra show good correspondence with experimental data.
\end{abstract}
\maketitle

\section{Introduction}
    Carbon nanotube (CNT) quantum devices have been the basis for diverse experimental and theoretical studies related to e.g.\ quantum information\cite{Churchill2009,Flensberg2010,Pei2012,Laird2013,Penfold-Fitch2017}, nano-electromechanical systems \cite{Steele2009,Lassagne2009,Benyamini2014}, induced \cite{Jarillo-Herrero2006} and artificially created\cite{Hamo2016} superconductivity, and predicted topological behavior\cite{Klinovaja2012,Sau2013,Marganska2018}. CNTs are attractive because their electronic behavior is well-understood and for sub-micron CNT based quantum dot devices, the electronic spectrum can be accurately described with a simple single-particle model. In this model each nearly four-fold degenerate longitudinal mode (shells)\cite{Liang2002,Cobden2002,Sapmaz2005} is described by valley ($\tau=K,K'$) and spin ($s=\uparrow, \downarrow$) degrees of freedom. Advances in fabrication techniques have led to high quality nanotube devices\cite{Cao2005,Waissman2013} which enable measurements of fine, spectroscopic features such as, e.g., spin-orbit interaction\cite{KuemmethNature2008,Churchill2009, JespersenNatPhys2011, Bulaev2008,Izumida2009, Klinovaja2011b} or disorder, which couple the bare quantum states in a well-defined manner. So far, the coupling of nanotube shells has not been examined in detail since the level spacing between shells in carbon nanotubes typically is so high that this coupling can be safely neglected.

    \begin{figure*}[htb!]
        \centering
        \includegraphics{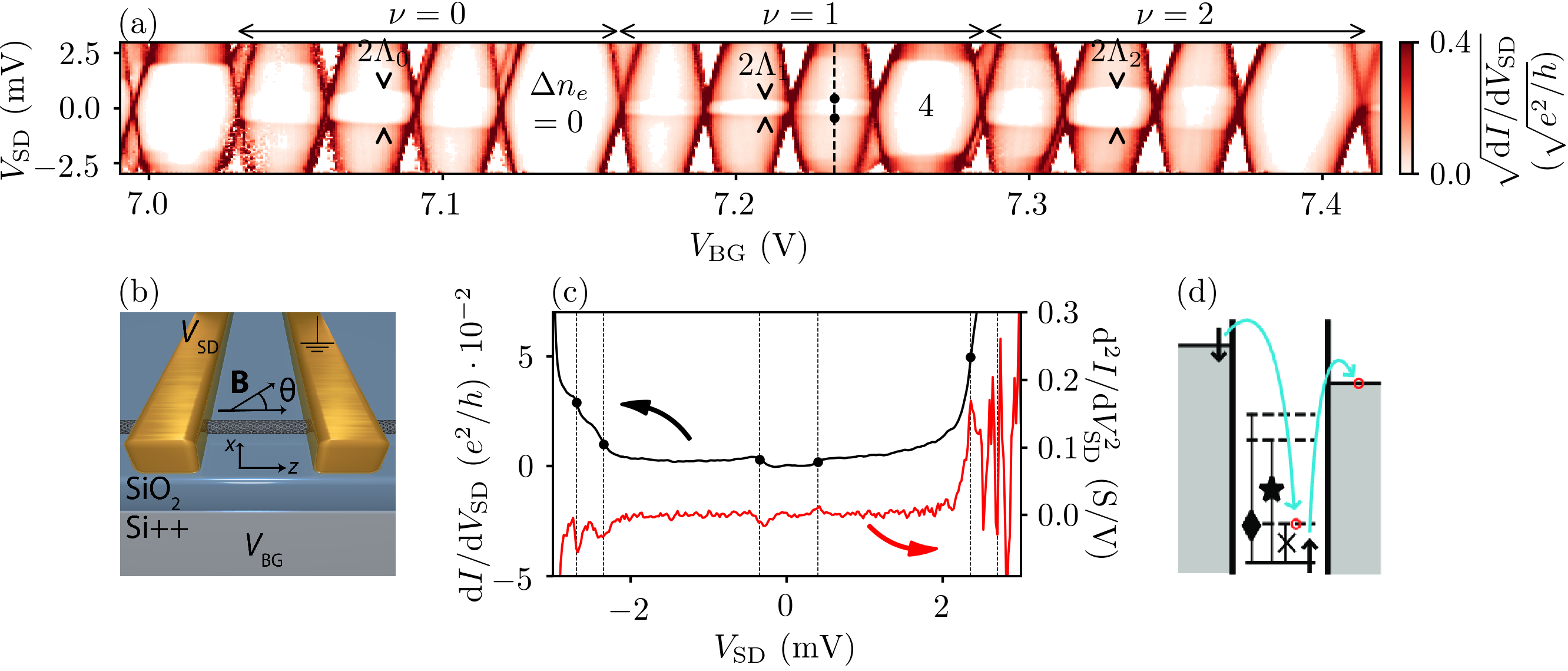}
        \caption{\label{fig:schematic+bias_spec}
            (a) Bias spectroscopy data in the conductance band showing conductance $\mathrm{d}I/\mathrm{d}V_{\text{SD}}$ as a function of applied bias $V_{\text{SD}}$ and backgate voltage $V_{\text{BG}}$. Diamond heights exhibit four-fold periodicity indicating filling of spin and valley degenerate shells whose indices are shown with $\nu$. The magnitude of the zero-field Kramers doublet splittings can be identified from the onset of inelastic cotunneling, highlighted as $\Lambda_{\nu}=\sqrt{(\Delta_{\text{SO}}^{\nu})^2+(\Delta_{KK'}^{\nu})^2}$.
            The colormap is shown as $\sqrt{\mathrm{d}I/\mathrm{d}V_{\text{SD}}}$ rather than $\mathrm{d}I/\mathrm{d}V_{\text{SD}}$ to emphasize these onsets whose conductance jumps are small relative to  sequential tunneling conductance.
            (b) Artist's representation of device (not to scale). The length of the quantum dot is defined by the separation between the electrodes which is 400~nm.
            (c) Cut through (a) at the red dashed line. The red (black) line shows conductance ($\mathrm{d}^2I/\mathrm{d}V_{\text{SD}}^2$). Steps in conductance correspond to peaks or dips in $\mathrm{d}^2I/\mathrm{d}V_{\text{SD}}^2$ and mark configurations where the applied bias $eV_{\text{SD}}$ is equal to the energy difference between two levels.
            (d) Schematic of inelastic cotunneling spectroscopy which is used to probe the spectrum of the nanotube quantum dot.
            Markers refer to Fig.\ \ref{fig:0e1e} at $5$~T.
        }
    \end{figure*}

    The first observations of the four-electron shell structure were reported in the early 2000s \cite{Liang2002, Cobden2002} followed by experiments establishing the near four-fold degenerate states as the starting point for more involved analysis of the observed carbon nanotube quantum states\cite{Minot2004,Cao2005, Jarillo-HerreroPRL2005, Jarillo-HerreroNature2005, Sapmaz2005, Maki2005, Makarovski:2006, SapmazSST2006, MakarovskiPRB2007, MakarovskiPRL2007, Grove-Rasmussen2007, Moriyama2007, Holm2008}. Initially the splitting of the four-fold degeneracy in two doublets was attributed to mixing of K and K' states (disorder), but the seminal experiment of Kuemmeth et al.\ in 2008 revealed that the spin-orbit coupling also plays a crucial role \cite{KuemmethNature2008,Churchill2009,Jhang2010,Ilani2010,SteeleNcomm2013, Izumida2009, Jeong2009,Chico2009,Logan2009}. Carbon nanotube quantum dots are typically analyzed within the single-particle model including spin-orbit coupling and disorder\cite{JespersenNatPhys2011, Schmid2015} even though interactions are shown to be important close to the band gap \cite{DeshpandeNPhys2008,Cleuziou2013,Pecker2013,LairdRMP2015, Niklas2016}.

    In this paper, we experimentally study the coupling of three shells in a CNT quantum
    dot and we extend the existing model to adequately include also inter-shell couplings\cite{Jarillo-HerreroNature2005,Grove2012,LairdRMP2015}, which allows for quantitative analysis of the data.
    The nanotube spectrum is probed experimentally with inelastic cotunneling spectroscopy \cite{Grove2012} which yields the transition energies between levels in the nanotube quantum dot.
    The evolution of these energy level transition energies is measured as a function of parallel, perpendicular and rotation of the magnetic field for various fillings of a nanotube shell.
    The quality of the model is assessed by calculating the excitation spectrum and fitting it to the obtained data.
    We find that the model fits the data well given two sets of parameters describing fillings of 0, 1 and 2, 3 and 4, respectively.

    \begin{figure}[ht!]
        \includegraphics[height=0.699\textheight]{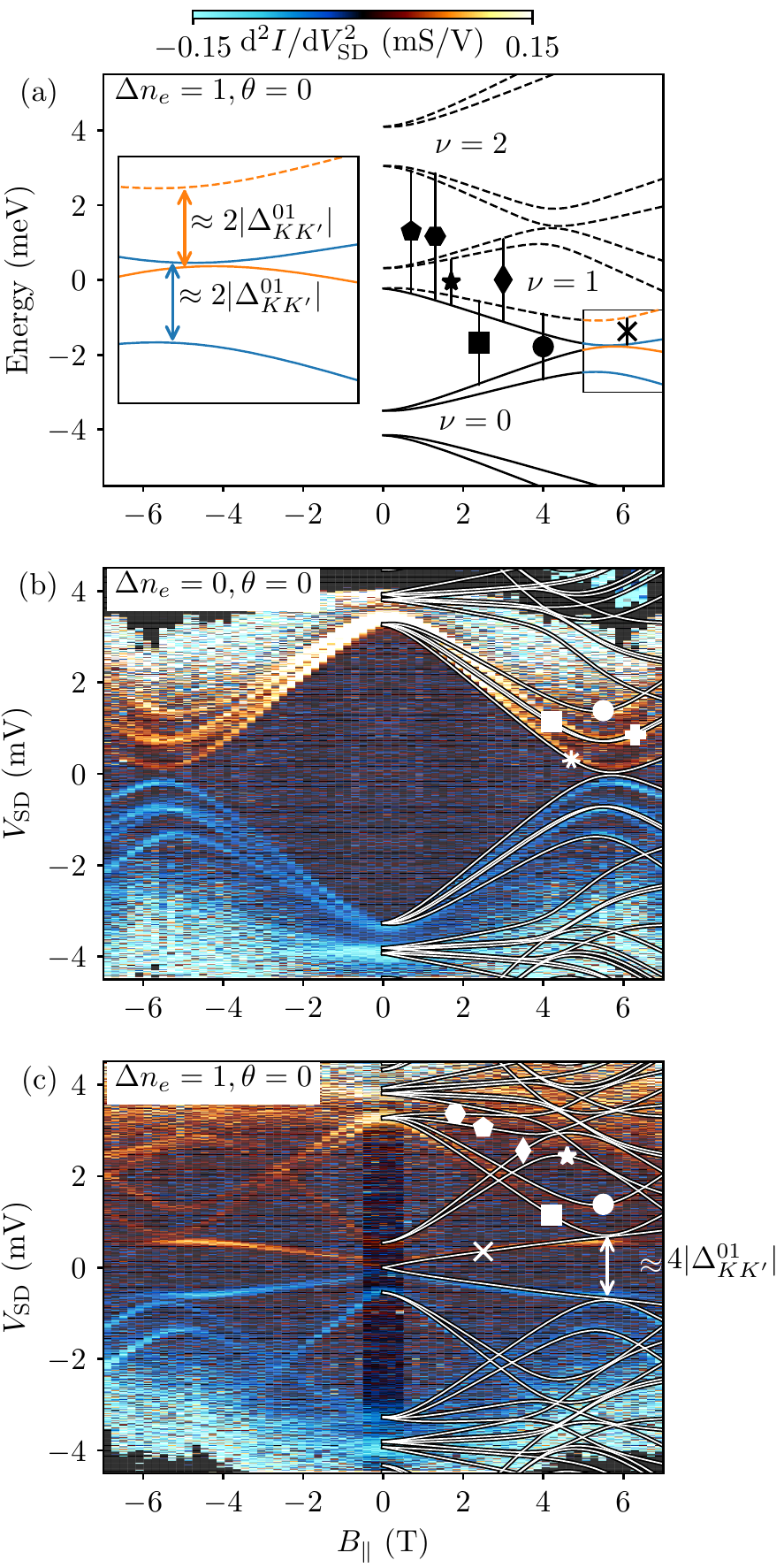}
        \caption{\label{fig:0e1e}
            (a) Spectrum of the three nanotube shells as a function of parallel magnetic field obtained from fitting experimental data in panel (b) and (c) to the model in Eq.\ \eqref{eq:model}.
            Solid (dashed) lines indicate filled (empty) states.
            The inset shows how the $\nu=0,1$ anticross magnitude (dashed square in panel (a)) depends on the inter-shell parameter $\Delta^{01}_{KK'}$.
            (b) and (c) Derivative of conductance $\mathrm{d}^2I/\mathrm{d}V_{\text{SD}}^2$ as a function of $V_{\text{SD}}$ and $B_{\parallel}$ in the center of Coulomb diamonds $\Delta n_e=0, 1$.
            The excitation spectrum is calculated from level differences in (a) and overlaid on the data.
            Excitations are identified by markers for easy comparison between model and data.
            Note that markers for high-energy excitations are left out for clarity.
        }
    \end{figure}

\section{Model}\label{sec:model}
    For the states in shell $\nu$ we will use an effective four-level model \cite{LairdRMP2015,JespersenNatPhys2011,Klinovaja2011b,Bulaev2008,Klinovaja2011a,Weiss2010} for a CNT quantum dot in an applied magnetic field with magnitude $B$ and angle $\theta$ measured from the nanotube axis:
    \begin{align}
        H_{\nu} =
        & g_{\text{s}}\mu_{\text{B}}B(\cos\theta\sigma_z\tau_0+\sin\theta\sigma_x\tau_0) \nonumber \\
        &+ g_{\text{orb}}^{\nu}\mu_{\text{B}}B\cos\theta\sigma_0\tau_z +
        \Delta_{\text{SO}}^{\nu}\sigma_z\tau_z
    \end{align}
    where $\tau_i$ and $\sigma_i$ are Pauli matrices in valley ($K$, $K'$) and spin space, $g_{\text{s}}$ the electron spin $g$-factor and $\mu_{\text{B}}$ the Bohr magneton.
    The effect of the magnetic field on the circumferential motion is opposite for $K$ and $K'$ and is parameterised by the orbital $g$-factor $g_{\text{orb}}^{\nu}$.
    $\Delta_{\text{SO}}^{\nu}$ sets the magnitude and sign of the spin-orbit interaction which couples spin and valley states.
    Each shell $\nu$ has its own set of parameters as indicated by the superscript.
    This is justified by experimental studies on separate shells which show that the parameters may change significantly between shells, but rarely change within a shell \cite{JespersenNatPhys2011,Hels2016}.

    Both shell index $\nu$, valley index $\tau$ and spin $s$ are conserved quantities in $H_{\nu}$ so we can label the eigenstates as $\ket{\nu\tau s}$.
    When imposing periodic boundary conditions around the circumference and hard-wall boundary conditions\cite{LairdRMP2015} at the nanotube-electrode interfaces we get the following wave functions for a metallic nanotube \cite{Bulaev2008,Weiss2010}
    \begin{align}
        \Psi_{\nu\tau s}(\phi,z) = \braket{\bvec{r}|\nu\tau s} = \frac{1}{\sqrt{\pi L}} e^{i\tau q\phi} \sin(\nu z\pi/L) \ket{s} .
    \end{align}
    Here $\nu=1,2,\ldots$, $\tau=\pm 1$ for $K,K'$.
    The nanotube quantum dot segment has length $L$, $\bvec{r}$ is the position vector for the electron, $z$ lies along the nanotube axis, and $\phi$ is along the circumferential direction.
    The orbital quantum number $q$ is defined by the chiral vector indices $n_1, n_2$ as $q=(n_1-n_2)/2$ which is an integer for metallic nanotubes.
    Note that the nanotube is only nominally metallic as it may still exhibit a (smaller) bandgap induced by curvature \cite{LairdRMP2015}.

    We now introduce a perturbation $H'$ to couple $K$ and $K'$ states motivated by disorder in the nanotube and interaction with the substrate
    \begin{align}
        H' = V(z)\delta(\phi) .
    \end{align}
    Here $V(z)$ is an atomically smooth perturbation in the longitudinal $z$ direction and $\delta(\phi)$ is an atomically sharp perturbation along the circumference.
    Note that $H'$ can only couple $K$ and $K'$ states if it contains an atomically sharp part \cite{LairdRMP2015}, and that this model does not consider e.g.\ the chirality of the CNT\cite{Izumida2015}.
    $H'$ leads to the following matrix elements
    \begin{align}
        \bra{\nu ms}H'\ket{\nu'm's'} = \Delta_{KK'}^{\nu\nu'}\delta_{ss'}
    \end{align}
    where
    \begin{align}\label{eq:Delta_def}
        \Delta_{KK'}^{\nu\nu'} = \frac{1}{\pi L} \int_0^L V(z)\sin(\nu z\pi/L)\sin(\nu'z\pi/L)dz .
    \end{align}
    Hence, this perturbation mixes all states in shell $\nu$ with all states in shells $\nu'$, except states with opposite spin.
    Note that Eq. \eqref{eq:Delta_def} implies $\Delta_{KK'}^{\nu\nu'}=\Delta_{KK'}^{\nu'\nu}$.

    For a constant $V(z)=V_0$ we obtain couplings within a shell ($\nu=\nu'$)
    \begin{align}\label{eq:structure}
        \bra{\nu ms}H'\ket{\nu m's'} = \Delta_{KK'}^{\nu} \equiv \Delta_{KK'}^{\nu\nu}\delta_{ss'} = \frac{V_0}{2}\sigma_0(\tau_0+\tau_x)
    \end{align}
    The $\tau_0$ term is often ignored when considering only a single shell because it simply amounts to a shift in energy which can be absorbed in the level spacings.
    The remaining $\tau_x$ describes the usual $KK'$ mixing.
    Here, we extend the standard model described above by allowing terms in the expansion of $V(z)$ which are first-order and above in $z$.
    These terms lead to the same structure as Eq. \eqref{eq:structure}, but they are off-diagonal in shell space.

    In the following we restrict ourselves to three shells labeled $\nu=0,1,2$ and separated by level spacings $\Delta E_{\nu\nu'}$, so that the full 12-dimensional Hamiltonian in $\nu$-space becomes
    \begin{align}\label{eq:model}
        H =&
        \begin{pmatrix}
            H_0 & 0 & 0 \\
            0 & H_1+\Delta E_{01} & 0 \\
            0 & 0 & H_2+\Delta E_{12}
        \end{pmatrix}
        + \Delta_{KK'}^{\nu\nu'}\delta_{ss'}
    \end{align}
    Each shell has three intrinsic parameters, $g_{\text{orb}}$, $\Delta_{\text{SO}}$ and $\Delta_{KK'}$, and there are three shell coupling parameters $\Delta_{KK'}^{\nu\nu'}$.
    This makes for a total of 14 independent parameters. Moreover, a shell can be described in terms of two Kramers doublets.
    Parameters or excitations that involve more than one shell (Kramers doublet) are termed inter-shell (inter-Kramers).
    Correspondingly, we use the term intra-shell (intra-Kramers) within a shell (Kramers doublet).

\section{Methods}
    Fig.\ \ref{fig:schematic+bias_spec}(b) shows the simple two-terminal geometry of the device.
    The nanotube is grown using chemical vapor deposition (CVD)\cite{Kong1998} on a doped Si substrate capped with a 500~nm capping layer of SiO$_2$.
    Subsequently, electrodes are defined with electron-beam lithography so that they are bridged by the nanotubes at random.
    The electrodes consist of Au/Pd (40/10~nm).

    Rotation of the magnetic field by angle $\theta$ in the $x$-$z$ plane was achieved using a piezo-electric rotator.
    Standard lock-in techniques were used to obtain $\mathrm{d}I/\mathrm{d}V_{\text{SD}}$.
    The lock-in conductance was differentiated numerically to obtain $\mathrm{d}^2I/\mathrm{d}V_{\text{SD}}^2$.
    Measurements were done at a temperature of 100~mK in a $^3$He/$^4$He dilution refrigerator.

    The CNT spectrum was probed with inelastic cotunneling spectroscopy to obtain the excitation spectrum.
    In this technique the applied voltage $V_{\text{SD}}$ is increased at a fixed magnetic field with the device in Coulomb blockade until it matches the energy difference between two levels.
    At this voltage a second-order tunneling process such as the one sketched in Fig.\ \ref{fig:schematic+bias_spec}(d) is allowed which causes an increase in conductance.
    Numerically finding the derivative of the conductance subsequently yields peaks whenever $e V_{\text{SD}}$ matches transition energies (see Fig.\ \ref{fig:schematic+bias_spec}(c)).

    The device measured in this paper has also provided data for previous studies \cite{JespersenNatPhys2011,Jespersen2011b}.

    \begin{figure*}[htb]
        \includegraphics{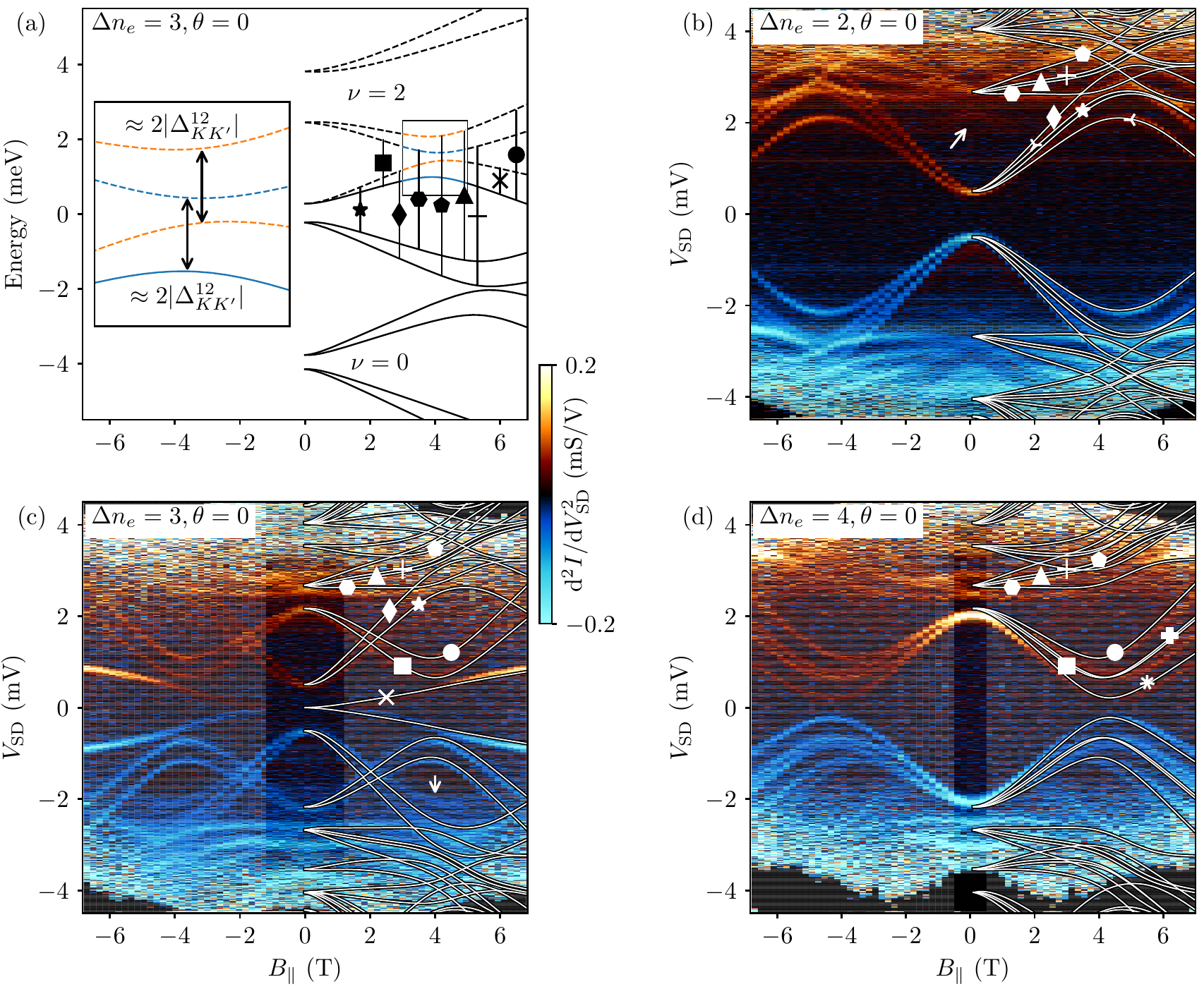}
        \caption{\label{fig:2e3e4e}
        Same as Fig.\ \ref{fig:0e1e} for $\Delta n_e=2, 3$ and $4$ in shell $1$.
        Note that in (a) only excitations between filled and empty states for a filling of 3 electrons are marked.
        This means that some marks in (b) and (d) are not found in (a).
        White arrows indicate excitations which are unexplained by the model (see text).
    }
    \end{figure*}

\section{Experimental Results}
    Initial characterization of the device using bias spectroscopy is shown in Fig.\ \ref{fig:schematic+bias_spec}(a).
    We plot $\sqrt{dI/dV}$ rather than $dI/dV$ to highlight the onset of inelastic cotunneling.
    The heights and widths of the Coulomb diamonds are seen to be approximately four-fold periodic, reflecting the filling of Kramers doublets in the nanotube shells. We label the electron filling of the dot by $\Delta n_e\equiv n_e - n_{e,0}$ and estimate an approximate occupation $n_{e,0}\approx 160$ electrons \cite{JespersenNatPhys2011}.
    At half-filling of shell $\nu$ ($\Delta n_e=-2,2,6$) the onset of inelastic cotunneling $\Lambda_{\nu}=\sqrt{(\Delta_{\text{SO}}^{\nu})^2+(\Delta_{KK'}^{\nu})^2}$ is marked on the figure by arrowheads.
    From the bias spectroscopy data we estimate the charging energies $U=7$--$8$~meV and level spacings $\Delta E=2$--$4$~meV.

    In order to investigate the shell couplings of the nanotube spectrum we perform inelastic cotunneling spectroscopy in shell $\nu=1$ for various fillings, magnetic field strengths $B$ and angles $\theta$ relative to the CNT axis.
    The model in Eq.\ \eqref{eq:model} is fitted to the data by manually iterating the parameters.
    The bias spectroscopy data in Fig.\ \ref{fig:schematic+bias_spec}(a) fixes some parameters and/or constrains the parameter space by providing $\Lambda_{\nu}$ and level spacings.
    Additionally, some intra-shell parameters are determined as in previous studies \cite{Hels2016} from data at low magnetic field where inter-shell couplings are negligible.
    Overall, we find parameter values consistent with those previously reported for similar devices \cite{KuemmethNature2008,Churchill2009,JespersenNatPhys2011,Lai2014,Cleuziou2013,Schmid2015}.
    Since $\Delta_{\text{SO}}\ll\Delta_{KK'}$ in all shells we can treat spin as an approximately good quantum number.
    This means that the two time-reversed states in a Kramers doublet have approximately opposite spin.

    \begin{table*}[htb]
        \caption{\label{tab:params}
            Parameters obtained from fitting inelastic cotunneling data in Figs.\ \ref{fig:0e1e}, \ref{fig:2e3e4e} and \ref{fig:2eperp_rot}.
            The parameters for 0 and 1 electrons in shell $1$ are different from those for 2, 3 and 4 electrons.
            This can be explained by a change in the electrostatic potential along the nanotube.
            All values are in meV except $g_{\text{orb}}$-values which are dimensionless.
            There is some uncertainty on the inter-shell parameters of $\approx 0.15$~meV which is correlated between the parameters.
            The uncertainty does not, however, affect the observation that two sets of parameters are needed.
            The intra-shell parameters $\Delta_{KK'}$ and $\Delta_{\text{SO}}$ ($g_{\text{orb}}$) have small uncertainties $\approx 0.05$~meV ($\approx 0.01$) since they are fixed by bias spectroscopy data (slopes of excitation spectroscopy data).
        }
        \begin{ruledtabular}
        \begin{tabular}{l|ccc|ccc|ccc|ccccc}
            shell &
            \multicolumn{3}{c|}{$\nu=0$} &
            \multicolumn{3}{c|}{$\nu=1$} &
            \multicolumn{3}{c|}{$\nu=2$} &
            \multicolumn{5}{c}{Inter-shell parameters} \\ \hline
            parameter &
            $\Delta_{\text{SO}}$ & $\Delta_{KK'}$ & $g_{\text{orb}}$ &
            \multicolumn{1}{c}{$\Delta_{\text{SO}}$} & $\Delta_{KK'}$ & $g_{\text{orb}}$ &
            \multicolumn{1}{c}{$\Delta_{\text{SO}}$} & $\Delta_{KK'}$ & $g_{\text{orb}}$ &
            \multicolumn{1}{c}{$\Delta E_{01}$} & $\Delta E_{12}$ & $\Delta_{KK'}^{01}$ & $\Delta_{KK'}^{12}$ & $\Delta_{KK'}^{02}$ \\ \hline
            $\Delta n_e=0, 1$     & 0.0 & 0.9 & $-6.4$ & 0.07 & 0.45 & $-5.5$ & 0.0 & 0.9 & $-8.7$ & 3.7 & 3.5    & 0.4    & 0.2     & 0.4 \\
            $\Delta n_e=2, 3, 4$  & 0.0 & 0.9 & $-6.4$ & 0.07 & 0.45 & $-6.2$ & 0.0 & 0.9 & $-6.2$ & 3.7 & 2.9    & 0.5    & 0.25    & 0.75 \\
            difference & $-$ & $-$ & $-$    & $-$  & $-$  & $-0.7$ & $-$ & $-$ & $+2.5$ & $-$ & $-0.6$ & $+0.1$ & $+0.05$ & $+0.35$ \\
        \end{tabular}
        \end{ruledtabular}
    \end{table*}

    For fillings $\Delta n_e=0, 1$ the obtained spectrum and data for parallel magnetic field ($B_\parallel$) are shown in Fig.\ \ref{fig:0e1e}.
    In the corresponding calculated spectrum for $\Delta n_e =1$ in Fig.\ \ref{fig:0e1e}(a), occupied (empty) energy levels are indicated by solid (dashed) lines.
    Excitations between occupied and empty levels are shown with vertical lines and a marker.
    Thus some excitations for, e.g., $\Delta n_e=0$ are not shown in Fig.\ \ref{fig:0e1e}(a) because they involve two filled or two empty levels.
    All three panels in Fig.\ \ref{fig:0e1e} share the same set of parameter values as listed in Table \ref{tab:params}.

    The experimental excitations in Fig.\ \ref{fig:0e1e}(b),(c) are all captured accurately by the model.
    At low magnetic field in Fig.\ \ref{fig:0e1e}(c) ($\Delta n_e=1$)
    the intra-Kramers excitation starts at zero energy due to the degeneracy at $B_\parallel=0$ and the two inter-Kramers excitations initially at $\Lambda_1$ split with approximately the electron $g$-factor.
    The fact that $\Delta_{\text{SO}}$ is non-zero is evident when comparing the lowest excitation in Fig.\ \ref{fig:0e1e}(c) ($\Delta n_e=1$) with the one in Fig.\ \ref{fig:2e3e4e}(c) ($\Delta n_e=3$).
    The former is convex while the latter is concave \cite{JespersenNatPhys2011}.

    Conversely, at low magnetic field no low-energy, intra-shell excitations are available for $\Delta n_e=0$ in Fig.\ \ref{fig:0e1e}(b) since all states in the $\nu=1$ shell are empty and the lowest excitation energy must therefore include a level spacing.
    By increasing the magnetic field the upper (lower) Kramers doublet in shell $\nu=0$ ($\nu=1$) are gradually brought closer until they anticross at $B_{\parallel}\approx 6$~T \cite{Jarillo-HerreroNature2005}.
    In Fig.\ \ref{fig:0e1e}(c) the same behavior for inter-shell excitations (square and circle) is observed. In fact, these excitations have the same energy in Fig.\ \ref{fig:0e1e}(b) as in (c) since adding one electron does not change these excitations.

    The anticross between shells $\nu=0$ and $\nu=1$ is shown in detail in the inset of Fig.\ \ref{fig:0e1e}(a).
    Blue levels anticross with blue, and orange with orange.
    Blue levels do \emph{not} anticross with orange levels since they have opposite spin.
     This prediction is confirmed by the data in Fig.\ \ref{fig:0e1e}(c) where the square and cross excitations do not repel each other to within the spectroscopic linewidth which is much smaller than the relevant inter-shell couplings $\Delta_{KK'}^{01}=0.4$~meV.
    The anticross magnitude is proportional to $|\Delta_{KK'}^{01}|$ as indicated by arrows.
    This magnitude is directly observable as $\approx 4|\Delta_{KK'}^{01}|$ in the data in Fig.\ \ref{fig:0e1e}(c) at $B_{\parallel}\approx 5.5$~T.
    Due to the finite spin-orbit coupling $\Delta_{\text{SO}}^{\nu=1}\neq 0$ the blue and orange states anticross at slightly different magnetic fields. Note, that for $n_e=0,1$, the $\nu = 1,2$ anticrosses are higher in energy and can not be resolved in the experiment.

    \begin{figure*}[htb]
        \centering
        \includegraphics{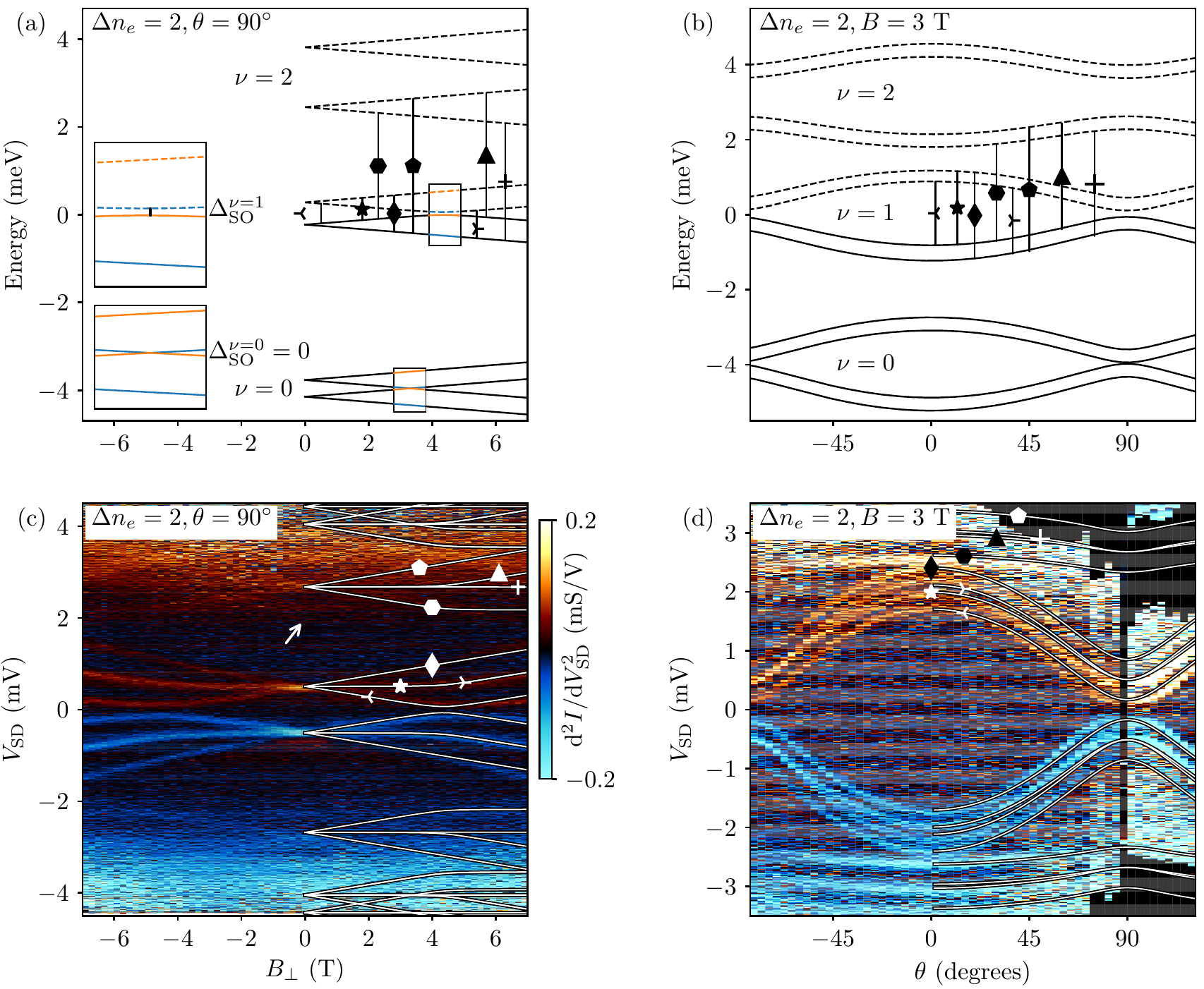}
        \caption{\label{fig:2eperp_rot}
            Same as Fig.\ \ref{fig:2e3e4e}, but for (a), (c) $\theta=90^{\circ}$ and (b), (d) magnetic field rotation.
            The insets in (a) show spin-orbit induced anticrosses, one of which is a crossing since $\Delta_{\text{SO}}^{\nu=0}=0$.
            These (anti)crossings are unrelated to shell couplings since they occur for states which belong to the same shell.
            In (b) no inset is shown since there are no ``simple'' single parameter anticrosses.
            The white arrow in (c) denotes the unexplained excitation which is also present in Fig.\ \ref{fig:2e3e4e}(b) (see text).
        }
    \end{figure*}

    To further investigate the excitations between the $\nu=1,2$ shells we repeat the procedure from Fig.\ \ref{fig:0e1e} for fillings $\Delta n_e=2, 3, 4$ in Fig.\ \ref{fig:2e3e4e}.
    Markers have been retained between Fig.\ \ref{fig:0e1e} and Fig.\ \ref{fig:2e3e4e} for the excitations that are present in both figures.
    The agreement between theory and the data is again excellent, although we find that some parameters must be adjusted for these new fillings to provide a good fit (see Table \ref{tab:params}). Almost all excitations visible in the data (Figs. \ref{fig:2e3e4e}(b)-(d)) are predicted quantitatively by the model with one set of parameters. As an example of a feature not resolved in Fig.\ 2, we identify the anticross for shell 1 and 2, illustrated in the inset of Fig.\ \ref{fig:2e3e4e}(a), at low bias in Fig.\ \ref{fig:2e3e4e}(c,d) with $\Delta_{KK'}^{12}=0.2$~meV.

    The parameters for Fig.\ \ref{fig:2e3e4e} are shown in Table\ \ref{tab:params} along with the difference in parameter values between the two sets of fillings $\Delta n_e=0,1$ and $\Delta n_e=2,3,4$.
    Most notable is the change in $g_{\text{orb}}^{\nu=2}$ of $+2.5$ and $\Delta_{KK'}^{02}$ of $+0.35$~meV.
    Adding electrons to the dot may change the electrostatic potential along the tube and this may explain the change in inter-shell parameters which are determined by $V(z)$.
    Theoretically, $g_{\text{orb}}$ is predicted to decrease with the number of electrons on the dot since the circumferential components of the constant Fermi velocity $v\sub{F}$ decreases as the longitudinal levels are filled\cite{Jespersen2011b}.
    Although $g_{\text{orb}}$ has been observed experimentally to vary with electron filling its dependence is not always systematic \cite{JespersenNatPhys2011,Hels2016}.
    The nanotube diameter is also predicted to influence $g_{\text{orb}}$ although independent measurements of diameter and $g_{\text{orb}}$ on the same nanotube are often inconsistent \cite{LairdRMP2015}.
    Overall, the variation of $g_{\text{orb}}$ is not understood.

    We note that if only the intra-shell excitations are considered, a single set of parameters is sufficient to describe all the data.
    As such, our results are consistent with previous studies on intra-shell excitations at low $B_{\parallel}$ field \cite{JespersenNatPhys2011,Hels2016} which found that the parameters did not change within a shell.

    Two features in the data in Fig.\ \ref{fig:2e3e4e} are unaccounted for in the model:
    At low magnetic field in Fig.\ \ref{fig:2e3e4e}(b) ($\Delta n_e=2$) at the white arrow a faint excitation is visible, gradually fading out above $B_{\parallel}=1$~T (also visible in Fig.\ \ref{fig:2eperp_rot}(c)).
    This excitation looks like the square and circle excitations from Figs.\ \ref{fig:2e3e4e}(c),(d) but it should not be present in the $\Delta n_e=2$ excitation spectrum since the corresponding states are empty.

    The second unexplained feature concerns the intra-Kramers excitations in Fig.\ \ref{fig:2e3e4e}(c) (diamond and asterisk).
    These arise from exciting an electron from occupied state in the lower Kramers doublet to the unoccupied state in the upper Kramers doublet.
    Thus, only two excitations are possible which is consistent with the data up to about $B_{\parallel}\approx 2.5$~T.
    Here, however, the degenerate excitations split in energy to reveal three excitations, the lowest of which (marked by a white arrow) is not captured in the model (this is most visible at negative $V_{\text{SD}}$) in Fig.\ \ref{fig:2e3e4e}(c).

    These qualitative inconsistencies can only be accounted for by a model which includes additional terms.
    For instance, including exchange interaction between shells 1 and 2 could induce a singlet-triplet splitting of the four-fold degenerate excitation above $V_{\text{SD}}\approx 2$~mV in Fig.\ \ref{fig:2e3e4e}(b) ($\Delta n_e=2$) which might explain the faint excitation at $V_{\text{SD}}\approx 2$~mV.

    To further verify the extracted parameters Fig.\ \ref{fig:2eperp_rot} shows excitation spectroscopy data for perpendicular orientation of the magnetic field (Fig.\ \ref{fig:2eperp_rot}(a),(c)) and rotation of the magnetic field (Fig.\ \ref{fig:2eperp_rot}(b),(d)), both for a filling of $\Delta n_e=2e$.
    Again, the calculated spectrum is superposed for $B_\perp > 0$ The parameters used are the same as in Fig.\ \ref{fig:2e3e4e} and the overall correspondence between data and theory is excellent, including the good correspondence of the $\blacklozenge$ excitation in Fig.\ \ref{fig:2eperp_rot}(c).
    This particular excitation involves two levels with approximately opposite spin so their separation is expected to increase proportional to $g\sub{s}$.
    Although $g\sub{s}$ is not a free parameter in the model the fit is still good.

    The splitting of the states in Fig.\ \ref{fig:2eperp_rot}(a),(c) is smaller than in previous figures since a perpendicular magnetic field does not couple to the orbital magnetic moment pointing along the CNT.
    Consequently, no shell anti-crossings are visible and we instead show intra-shell anti-crossings caused by $\Delta_{\text{SO}}$.
    In the model, the intra-shell spin-orbit coupling for shell $\nu=0$ is set to zero since the data needed to estimate $\Delta_{\text{SO}}^{\nu=0}$ is not available (see resulting level crossing in lower inset of Fig.\ \ref{fig:2eperp_rot}(a)).

    In Fig.\ \ref{fig:2eperp_rot}(d) the fact that the \markerfour\ and \markerasterisk\ excitations have a finite splitting in parallel field and no splitting in perpendicular field is another indication of the finite spin-orbit coupling \cite{JespersenNatPhys2011}.
    At perpendicular field (see Fig.\ \ref{fig:2eperp_rot}(a)) the orbital motion does not couple to the magnetic field.
    The resulting energy levels are split purely by spin, leading to particle-hole symmetry and consequently to degenerate excitations.
    Conversely, at parallel magnetic field (Fig.\ \ref{fig:2e3e4e}(a)), spin-orbit interaction causes a slight asymmetry between the upper and lower Kramers doublet and a corresponding splitting (different magnitude) of the \markerfour\ and \markerasterisk\ excitations, which is clearly observed in the data.

\section{Conclusion}
    We have studied experimentally and theoretically the couplings and excitations between three shells in a carbon nanotube quantum dot.
    The results show that the magnetic field behavior of the energy levels of three shells can be accurately captured by extending an existing shell model.
    However, contrary to expectations, we find that the parameters $g_{\text{orb}}^{\nu=1}$, $g_{\text{orb}}^{\nu=2}$, $\Delta E_{12}$, $\Delta_{KK'}^{01}$, $\Delta_{KK'}^{12}$, and $\Delta_{KK'}^{02}$ change when adding the second electron to one of the considered shells. The change in inter-shell parameters may be due to a change in the electrostatic potential caused by the added electron, while the change in $g_{\text{orb}}$ currently not understood.
    Finally, the clear identification of disorder and intrinsic spin-orbit induced anti-crossings in the level structure constitute a valuable reference for future studies. In particular, artificially created spin-orbit coupling by electric fields\cite{Klinovaja2011} or micromagnet pattering \cite{Kjaergaard2012} may lead to additional intershell couplings, which can be probed by carbon nanotube quantum dot bias spectroscopy.

\begin{acknowledgments}
    We would like to thank Bernd Braunecker, Jens Paaske and Karsten Flensberg for fruitful discussions and acknowledge the financial support from the Carlsberg Foundation, Villum foundation, the European Commission FP7 project SE2ND, the Danish Research Councils and the Danish National Research Foundation.
\end{acknowledgments}

%


\end{document}